\documentclass[12pt]{article}
\input psfig.sty
\usepackage{natbib}

\def\refitem{\noindent\hangindent 20pt}
\def\wisk#1{\ifmmode{#1}\else{$#1$}\fi}
\def\lt     {\wisk{<}}

\def\lsim   {\wisk{_<\atop^{\sim}}}

\def\deg    {\wisk{^\circ}}

\begin{document}
\normalsize

\renewcommand{\bottomfraction}{0.7}

\begin{center}
{\Large Reionization and Structure Formation with ARCADE}
\end{center}

\begin{center}
A. Kogut\footnote{Code 685, Goddard Space Flight Center, Greenbelt, MD 20771}
\end{center}

\noindent
Structure formation in the early universe
is expected to produce a cosmological background of free-free emission.
The Absolute Radiometer for Cosmology, Astrophysics, 
and Diffuse Emission (ARCADE)
will measure the spectrum of the cosmic microwave background
at centimeter wavelengths
to detect this signature of structure formation.
I describe the ARCADE instrument
and the cryogenic engineering required
to achieve mK accuracy at such long wavelengths.

\section{Introduction}

In standard cosmological models,
the cosmic microwave background (CMB)
originates in a hot, dense phase of the early universe.
Matter and radiation remain tightly coupled 
as the universe expands and cools,
until the temperature drops sufficiently below the ionization potential
of hydrogen at redshift $z \sim 1100$.
Structures then emerge through gravitational infall and collapse,
until the first collapsed objects ignite in nuclear fusion
to re-ionize the intergalactic medium and
end the cosmic ``dark ages.''

The non-linear collapse of bound structures
depends sensitively on poorly known parameters
such as the distribution function of rare high-density peaks,
the efficiency of competing cooling mechanisms,
and the disruptive ``feedback'' of ionizing radiation
from the first objects on successive generations
(see, e.g., 
Tegmark et al.\ 1997,
Gnedin \& Ostriker 1997,
Haiman \& Loeb 1997,
Oh 2001).
Measurements of the
Gunn-Peterson absorption trough 
in spectra of distant quasars
show neutral hydrogen present at $z \approx 6$
(Becker et al.\ 2001,
Djorgovski et al.\ 2001,
Fan et al.\ 2002).
WMAP measurements of correlated temperature-polarization anisotropy,
however,
show a substantial optical depth to free electrons 
at much higher redshifts,
indicating that reionization occurred 
within the redshift range $11 < z_r < 30$
(Kogut et al.\ 2003).
The two results are not incompatible --
since absorption spectra are sensitive 
to even small amounts of neutral hydrogen,
models with partial ionization $x_e \lsim 1$
can have enough neutral column density
to produce the Gunn-Peterson trough
while still providing free electrons
to scatter CMB photons and produce large-scale polarization.

Spectral measurements at long wavelengths
provide additional information
on reionization and structure formation.
The ionized intergalactic medium 
can cool through free-free emission
from electron-ion interactions.
The distortion to the present-day CMB spectrum is given by
\begin{equation}
\Delta T_{\rm ff} = T_{\gamma} \frac{Y_{\rm ff}}{x^2}
\label{FF_distortion_eq}
\end{equation}
where $T_{\gamma}$ is the undistorted photon temperature,
$x$ is the dimensionless frequency $h \nu / k T_{\gamma}$,
and
$Y_{\rm ff}$ is the optical depth to free-free emission
(Bartlett \& Stebbins 1991).  
The distortion from a blackbody spectrum
varies as $\nu^{-2}$
and can be appreciable at sufficiently long wavelengths.

\begin{figure}[b]
\centerline{
\psfig{file=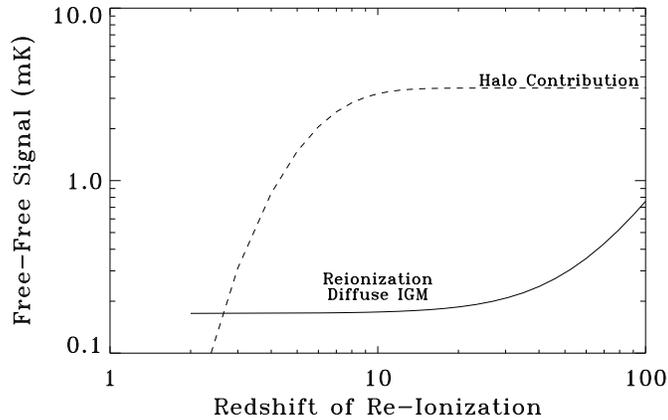,height=2.5in,angle=90}}
\caption{
Cosmological free-free signal at 2 GHz
as a function of the redshift of reionization.
The contribution from clumpy halos at $z \sim 10$
dominates the more smoothly distributed signal from reionization
at higher redshifts.
}
\label{ff_vs_z}
\end{figure}

Figure \ref{ff_vs_z}
illustrates the dependence of cosmological free-free emission
on reionization and structure formation.
The amplitude of the cosmological free-free signal 
depends on the column $\int n_e^2$ of ionized gas
and thus on the redshift $z_r$ 
at which the first collapsed objects formed.
Using the observed Lyman-$\alpha$ forest,
Haiman \& Loeb (1997)
derive a lower limit
$|Y_{\rm ff}| > 8 \times 10^{-8}$ 
for the cosmological free-free background
originating at redshift $z < 5$,
corresponding to a distortion
$\Delta T = 0.2$ mK
at 2 GHz.
Ionized gas at higher redshift will further enhance the signal.
Two sources are important.
Reionization produces a smoothly-distributed signal at high redshift,
while a more clumpy component from halos
occurs at intermediate redshift $z \sim 10$.
Since $\Delta T \sim n_e^2$,
the integrated spectral distortion
is strongly weighted to denser regions.
Oh (1999) shows that
the halo contribution is dominant,
with mean distortion $\Delta T \approx 3$ mK at 2 GHz.
Measurements at long wavelengths
thus probe the thermal history of the universe
in the redshift range 10--30
where the bulk of structure formation occurs.

\begin{figure}[t]
\centerline{
\psfig{file=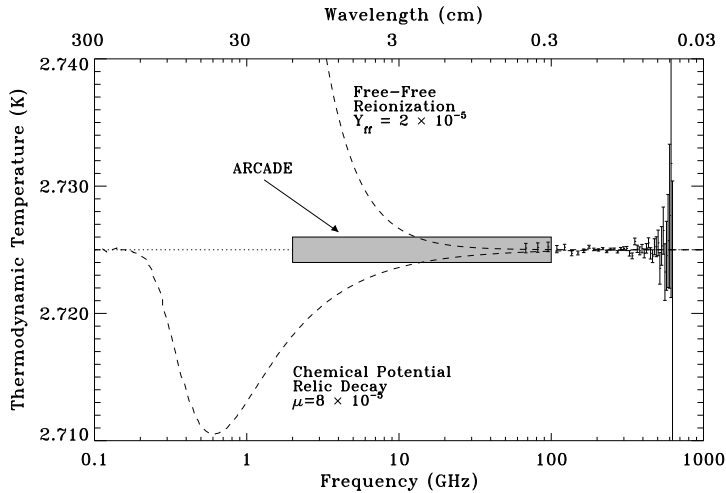,height=2.5in,angle=90}}
\caption{
Current 95\% confidence upper limits to distorted CMB spectra.
The FIRAS data and ARCADE 1 mK error box are also shown;
error bars from existing cm-wavelength measurements are larger than
the figure height.
Free-free emission from reionization and structure formation
causes a quadratic increase in temperature at long wavelengths.
Energy release at very high redshift ($z > 10^3$)
could also create a chemical potential distortion.
}
\label{arcade_vs_firas}
\end{figure}

Figure \ref{arcade_vs_firas}
shows current upper limits 
to CMB spectral distortions.
Direct observational limits at long wavelengths are weak:
distortions as large as 5\% could exist 
at wavelengths of several centimeters or longer
without violating existing observations.
Structure formation
should produce a cosmological free-free background
with amplitude of a few mK at frequencies of a few GHz.
Such a signal is well below current observational limits,
which only constrain
$|Y_{\rm ff}| ~\lt ~1.9 \times 10^{-5}$
(Bersanelli et al.\ 1994).

Long-wavelength measurements of the CMB spectrum 
are also sensitive to distortions created
at earlier epochs ($10^4 < z < 10^7$), 
described by a Bose-Einstein distribution
with dimensionless chemical potential 
$
\mu = 1.4 ~ \Delta {\rm E} / {\rm E},
$
proportional to the fractional energy release to the CMB
(Burigana et al.\ 1991).
The resulting spectrum has a distinctive drop in brightness temperature
at centimeter wavelengths (Figure \ref{arcade_vs_firas}).
A chemical potential distortion is a primary signature for
the decay of relics from GUT and Planck-era physics.

\section{ARCADE Instrument}

\begin{figure}[b]
\vbox to2.8in{\rule{0pt}{2.8in}}
\includegraphics{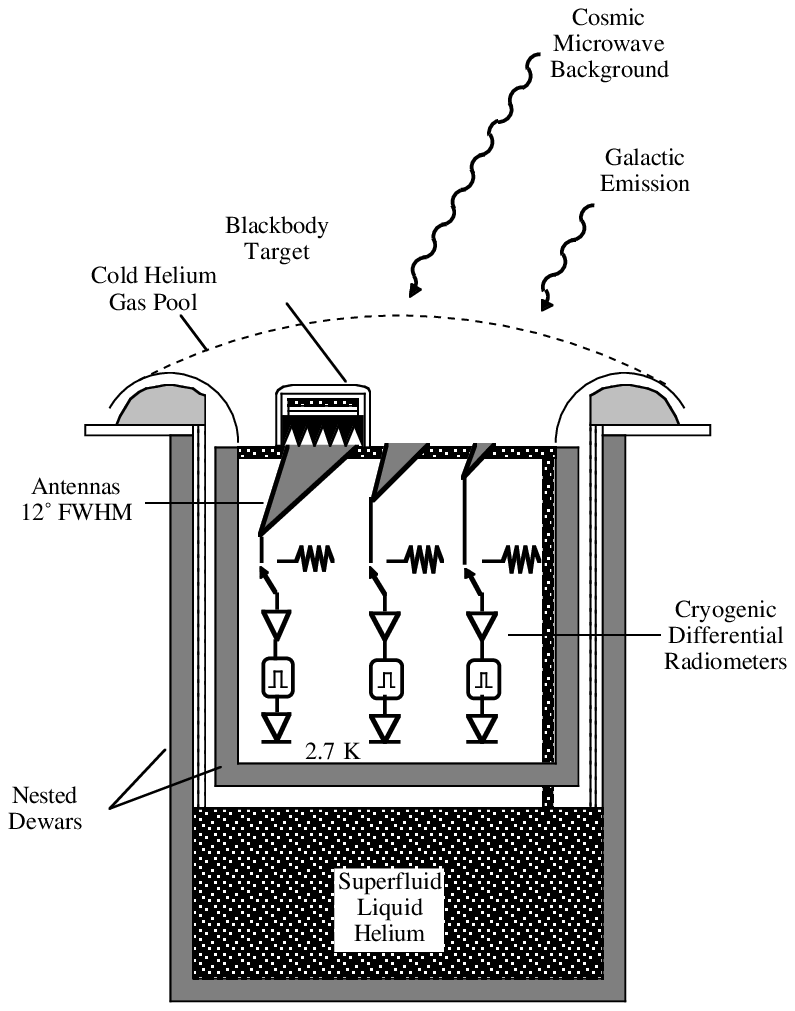}
\includegraphics{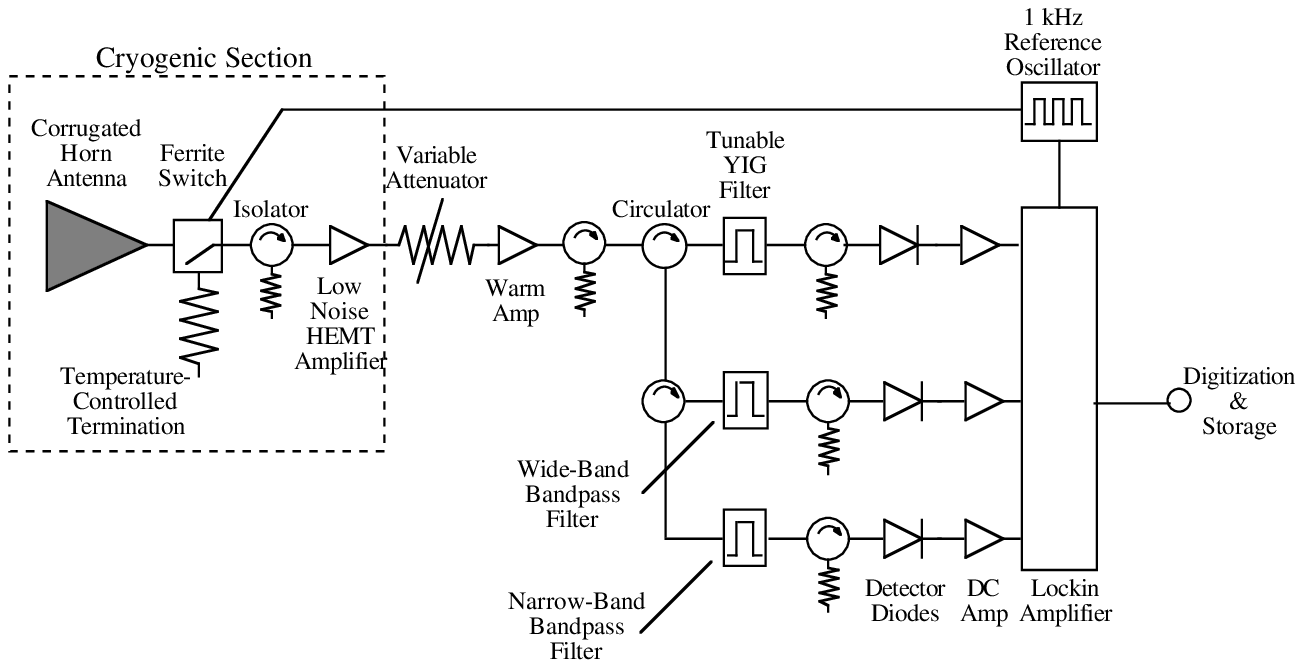}
\caption{
(Left) Schematic drawing of ARCADE instrument.
(Right) Block diagram of ARCADE radiometers.}
\label{arcade_schematic}
\end{figure}

The Absolute Radiometer for Cosmology, Astrophysics, and Diffuse Emission
(ARCADE)
is an instrument designed to detect the cosmological signal from
reionization and structure formation
through long-wavelengh measurements of the CMB spectrum.
Figure \ref{arcade_schematic} shows the instrument design.
It consists of 6 balloon-borne narrow-band cryogenic radiometers
($\Delta \nu / \nu \sim 10\%$)
with central frequencies $\nu$ = 3.3, 5.5, 8.1, 10.1, 30.3, and 89 GHz
chosen to cover
the gap between full-sky surveys at radio frequencies ($\nu < 2$ GHz)
and the FIRAS millimeter and sub-mm measurements.
Each radiometer measures the difference in power
between a beam-defining antenna (FWHM $\sim 12\deg$)
and a temperature-controlled internal reference load.
An independently controlled blackbody target 
(emissivity $\epsilon > 0.9999$) is located on the aperture plane,
and is rotated to cover each antenna in turn,
so that each antenna alternately views the sky 
or a known blackbody.

The ARCADE instrument design greatly reduces all
major sources of systematic error 
in previous long-wavelength spectral measurements.
It is a differential instrument with an external blackbody target:
instrumental signals thus drop out in the sky--target comparison.
It observes above the bulk of the atmosphere,
reducing atmospheric emission to below 2 mK.
The instrument is fully cryogenic;
all major components are independently temperature-controlled 
to remain at 2.7 K, isothermal with the signal from deep space.
Boiloff helium vapor,
vented through the aperture plane,
forms a barrier between the instrument and the atmosphere;
there are no windows or warm optics to correct.
All channels observe a common external calibration target,
eliminating cross-calibration uncertainties.
The experimental precision is limited 
only by systematic uncertainties
associated with operation within the Earth's atmosphere
and is intended in part
to develop technology for the DIMES space mission.

Each radiometer consists of a cryogenic front end
with a high electron mobility transistor (HEMT) 
direct-gain receiver,
switched at 1 kHz
for gain stability
between a wavelength-scaled corrugated conical horn antenna
and a temperature-controlled internal load.
To reduce insertion loss
and the resulting sensitivity to instrument temperature,
we switch between the sky horn and internal load
using latching ferrite switches
whose measured insertion loss is below 0.4 dB.
The low insertion loss greatly reduces requirements for
thermal monitoring and control of the instrument front end.
The back end of each radiometer is split into two frequency sub-channels:
a wide-band channel for maximum sensitivity,
and a narrow-band channel restricted to protected RF bands.
A lockin amplifier in each band demodulates the switched signal
to produce an output proportional to the difference in power
between the antenna and the internal load.
All of the radiometer back ends are housed 
in a temperature-controlled module mounted to the outside of the dewar.

Each receiver is fed by a corrugated conical horn antenna
scaled in wavelength to produce identical beam shape
in each frequency channel.
To avoid convective instabilities in the helium vapor barrier,
the instrument should remain vertical during observations.
We reconcile this requirement 
with the need for sky coverage (and avoiding direct view of the balloon)
by mounting the antennas at a 30\deg ~angle from the zenith,
slicing each antenna at the aperture plane.
Quarter-wave chokes surround the resulting elliptical aperture
and provide further suppression of side lobes.
The cold flares at the dewar rim function as a cold ground shield
to block emission from the Earth.  

The external target consists of a microwave absorber
(Eccosorb CR-112, an iron-loaded epoxy)
cast with grooves in the front surface to reduce reflections.
The Eccosorb is mounted on a series of thermally conductive 
plates with conductance $G_1$
separated by thermal insulators of conductance $G_2$.
Thermal control is achieved by heating the outermost buffer plate,
which is in weak thermal contact with a superfluid helium reservoir.
Radial thermal gradients at each stage 
are reduced by the ratio $G_2 / G_1$ between the buffer plates.
Fiberglass and copper achieve 
a ratio $G_2 / G_1 \lt 10^{-3}$;
a two-stage design should achieve net thermal gradients 
well below 1 mK,
verified both in the lab and in flight using
embedded ruthenium oxide thermistors.
We calibrate the thermistors {\it in situ}
by immersing the target in a LHe bath,
then recording data through the flight thermal readout system
as the pressure above the bath is varied.
This calibrates out residual (nW) self-heating in each thermometer
while providing accurate
cross-calibration between thermometers
to limit thermal gradients in flight.
The readout system has demonstrated in-flight 
white noise below 0.2 mK Hz$^{-1/2}$
(Fixsen et al.\ 2002).
We have measured the target power reflection coefficient
with the target in place over the flight antennas.
The resulting upper limits $R < -35$ dB
exceed the -30 dB design requirement.

ARCADE's observing strategy is straightforward.
Independent thermal control in each channel
adjusts the internal load temperatures
to null the radiometer outputs
when the antennas view the sky.
We than adjust the external target temperature
to exactly null the sky signal in the longest-wavelength channel.
With all temperatures held constant,
the target will then move to cover the short-wavelength antennas.
If the sky were a perfect blackbody,
all shorter-wavelength channels
would also show a null
between the sky and target.
ARCADE will thus measure small spectral shifts about a precise blackbody,
greatly reducing dependence on instrument calibration and stability.
By comparing each channel to the {\it same} external target,
uncertainties in the target temperature cancel
so that deviations from a blackbody spectral {\it shape}
may be determined much more precisely than the absolute temperature.
The ARCADE nulling scheme allows more than an order of magnitude
improvement over previous cm-wavelength measurements,
and is directly analogous to the nulling used to achieve
50 parts-per-million accuracy by the FIRAS experiment
(Mather et al. 1999).

\section{Stupid Dewar Tricks}

\begin{figure}[b]
\vbox to2.5in{\rule{0pt}{2.5in}}
\includegraphics{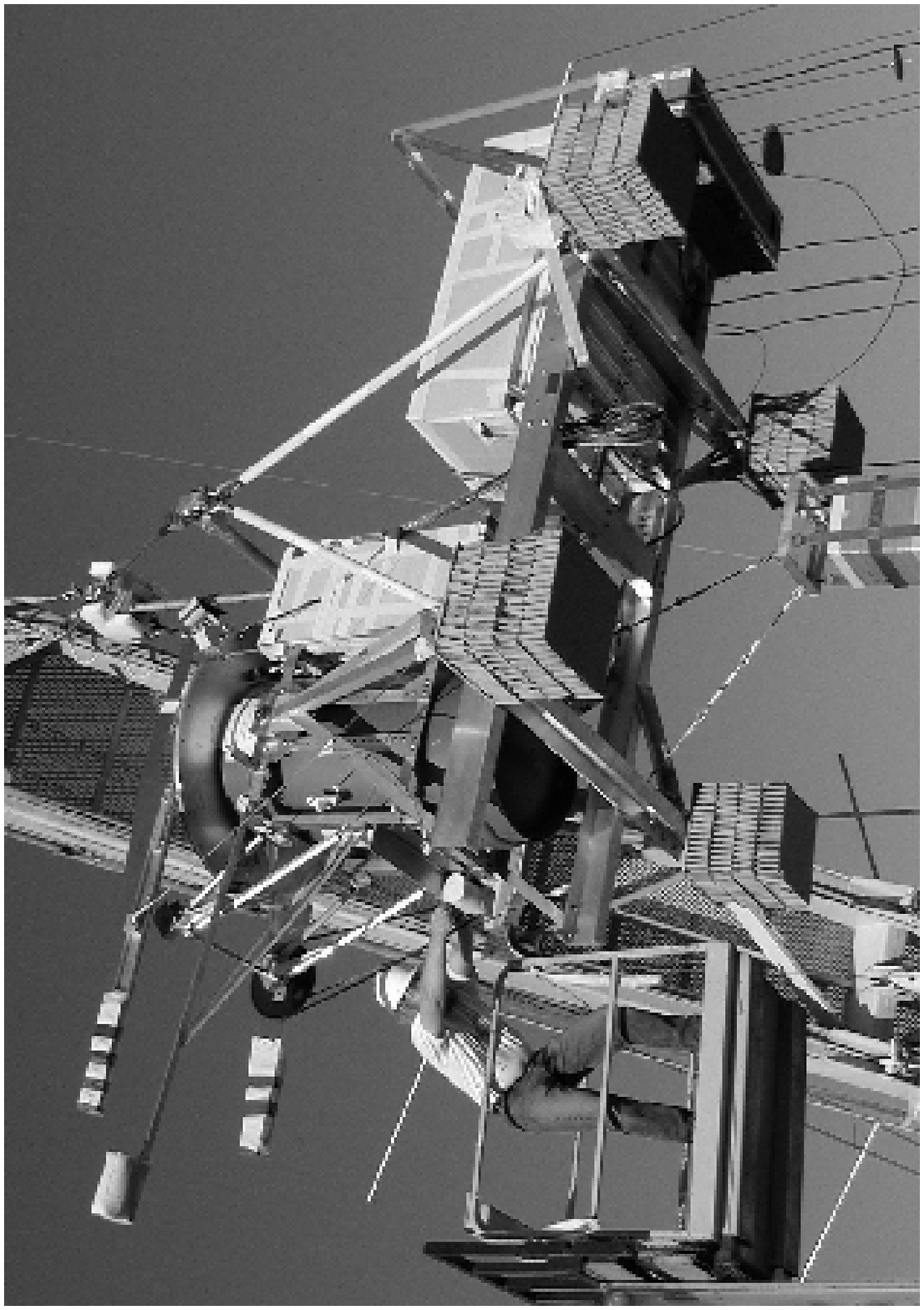}
\includegraphics{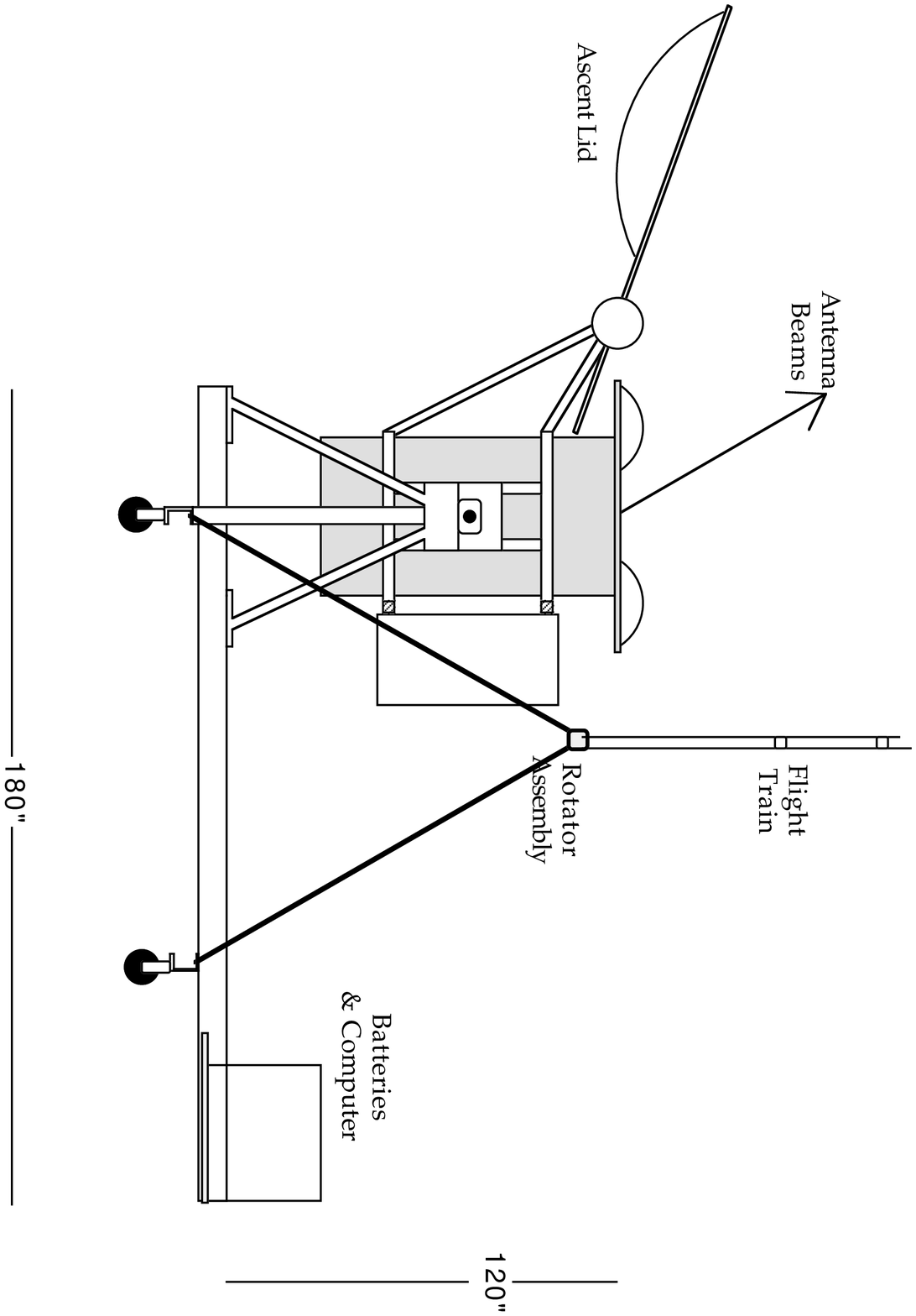}
\caption{
(left) Payload in 2001 configuration
with 2-channel prototype instrument.
(right) Gondola schematic for 2001 flight.
The dewar is counter-balanced by the batteries and flight electronics.
}
\label{two_gondolas}
\end{figure}

Figure \ref{two_gondolas}
shows the gondola design for the 2001 flight.
It consists of a 3.7 m frame
with the dewar on one end.
Batteries and flight electronics 
serve as counterweights on the other end,
allowing the suspension to be located well away from the dewar.

The ARCADE design has a fully cryogenic clear aperture
operating without windows
or any view to warm components.
Achieving this in practice requires
some innovative crogenic engineering.
Normally, one might isolate the instrument from 
ambient (250 K) conditions
by putting the cryogenic components
toward the bottom of the dewar.
But this would entail some view of the dewar wall,
which necessarily has a large temperature drop.
In order to eliminate any emission from the dewar walls,
the ARCADE aperture plane is located 
at the {\it mouth} of an open bucket dewar,
using a second (internal) dewar to reduce radial heat transport
between the 2.7 K optics and the 250K exterior wall some 20 cm away.
To keep the aperture cold,
ARCADE uses fountain-effect pumps
to move superfluid LHe from storage in the bottom of the dewar
to reservoirs located in the aperture plane 
and external calibration target.
Boiloff gas passes through pinholes 
in the aperture plane, 
providing additional cooling through the enthalpy of the gas.

ARCADE operates at altitudes above 33 km.
An ascent lid covers the aperture during launch,
then opens to allow observations of the sky.
Once the lid is open, 
we prevent condensation of atmospheric nitrogen onto the cold optics
using boiloff He gas.
Helium gas at ambient pressure of 3 Torr
is denser than atmospheric nitrogen
provided the helium remains colder than 20 K.
Stainless steel flares, also cooled by boiloff gas,
surround the cryogenic aperture 
to trap a pool of cold He gas over the optics.
Resistive heaters in the LHe reservoir
provide control over the He gas flow,
which can be increased to maintain cold enough temperatures
above the aperture plane.

\begin{figure}[b]
\vbox to2.2in{\rule{0pt}{2.0in}}
\includegraphics{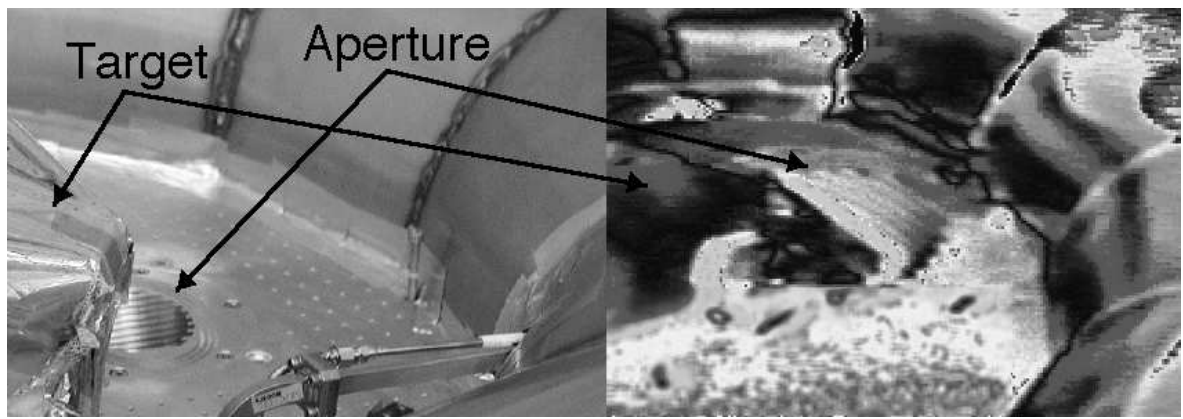}
\caption{
Aperture plane with target and antenna aperture.
(left) 30 GHz aperture during ground testing.  
The target is visible in the left foreground;
the cold flares are in the background.
(right) Video still taken in flight,
30 minutes after opening ascent lid.
The corrugations of the 10 GHz aperture
are visible behind the target in the foreground --
nitrogen condensation is not a major problem.}
\label{video_still}
\end{figure}

We have successfully flown a 2-channel prototype
to validate the open-aperture cryogenic design,
demonstrating the ability to control the 
radiometer front end and cold optics at 2.7 K.
Figure \ref{video_still}
shows an image of the aperture plane
taken during the November 2001 flight.
More than 30 minutes after opening the lid,
only a few mm of condensation has formed.
Thermal data from the same flight are consistent
with slow ice accumulation of 0.1 g/s.
Nitrogen condensation did {\it not} begin
immediately after the ascent lid opened,
but was apparently triggered 5 minutes later
by turbulent mixing above the aperture plane.
Despite the slow buildup,
we were able to obtain over 30 minutes of good data on the sky,
sufficient to measure the CMB spectrum.
Since the instrument is nearly isothermal with the CMB
and nitrogen ice is nearly transparent to microwaves,
minor condensation on the flight optics
will not cause significant effects.
Following the November 2001 flight, 
we have modified the aperture plane
to decrease turbulent mixing
and further reduce nitrogen condensation.
A kW heater on the aperture plane
now provides the ability to remove accumulated ice
several times per flight.

Mechanical failure of the external target position control
limited science returns from the first flight.
A second flight is scheduled for summer 2003
to test the modified aperture plane,
including the de-icing heater.
This second flight will 
compare the CMB temperature between 10 and 30 GHz
to accuracy of a few mK.
A larger version using all 6 channels
is under construction
and is expected to fly before 2005.

\section{Error Budget}

ARCADE will not be limited by raw sensitivity.
HEMT amplifiers cooled to 2.7 K
easily achieve {\it rms} noise $\sim$1 mK Hz$^{-1/2}$:
a few seconds integration provides all the sensitivity required.
We verify that the 1/f knee is below 
the corresponding frequency of 0.03 Hz.
Table \ref{error_budget} summarizes the ARCADE error budget.
ARCADE is designed to reduce, eliminate, or control 
the major sources of systematic uncertainty
which have limited previous spectral measurements at centimeter wavelengths.
The ARCADE spectra are derived from comparison of the sky 
to the external blackbody target
in multiple frequency channels,
and are thus sensitive only to the
{\it difference} of systematic effects from channel to channel.

\begin{table}[t]
\caption{\label{error_budget}
ARCADE Error Budget}
\begin{center}
\begin{tabular}{l c r r r}
\hline
Source & Measurement & \multicolumn{3}{c}{Amplitude and Uncertainty (mK)} \\
       & Technique   & 3 GHz & 10 GHz & 90 GHz \\
\hline
Earth 		  & Beam, Tip Scan	
	& $1 \pm 0.5$   & $0.3 \pm 0.2$ & $0.3 \pm 0.2$ \\
Flare Emission	  & Beam, Thermistors	
	& $1.5 \pm 0.5$ & $0.5 \pm 0.2$ & $0.5 \pm 0.2$ \\
Flight Train	  & Tip Scan, Screen
	& $3.6 \pm 1.2$ & $3.6 \pm 1.2$ & $3.6 \pm 1.2$ \\
Balloon		  & Beam, Tip Scan	 
	& $0.2 \pm 0.1$ & $1.7 \pm 0.5$ & $3.6 \pm 1.8$ \\
Target Gradients  & Thermistors		
	& $1.0 \pm 0.5$ & $0.3 \pm 0.2$ & $0.3 \pm 0.2$ \\
Target Drifts	  & Thermistors		
	& $0.2 \pm 0.2$ & $0.2 \pm 0.2$ & $0.2 \pm 0.2$ \\
Atmosphere	  & Altitude, Tip Scan
	& $1.4 \pm 0.05$ & $1.6 \pm 0.06$ & $9.8 \pm 0.4$ \\
Target Reflection   & Lab 		
	& $\lt 0.1$ & $\lt 0.1$ & $\lt 0.1$  \\
Calibration	  & Target Temperature  
	& $\lt 0.1$ & $\lt 0.1$ & $\lt 0.1$  \\
Noise		& Lab, Binning
	& $\lt 0.1$ & $\lt 0.1$ & $\lt 0.1$  \\
\hline
\multicolumn{2}{l}{Quadrature Uncertainty}
	& $\pm 1.5$ & $\pm 1.4$ & $\pm 2.2$ \\
\hline
\end{tabular}
\end{center}
\end{table}

The largest systematic uncertainties arise from
emission of warm objects outside the dewar
(the flight train and Earth).
We estimate a worst-case signal from the flight train using
the gondola geometry and measured beam patterns.
The rotator, flight train, and suspension cables will be hidden
behind reflectors to redirect the beam to blank areas of the sky.
The signal from the reflectors is dominated by their emission
and by stray glints to the ground.
The largest signals are from the atmosphere,
balloon, and flight train.
The flight train is situated at nearly the same relative geometry
with respect to all antennas,
partially cancelling this signal in the differential sky spectra.
We will further reduce the uncertainty by
using tip scans and azimuthal rotation to
modulate the signal from the flight train.
Ground screens may provide still further reduction.
The combination of differential measurement,
tip scans, rotational modulation,
and screens
will reduce the uncertainty in flight train emission or reflection
by a factor of 2--3.
A similar exercise provides an in-flight estimate for reflection
from the balloon;
since the balloon is nearly transparent below 10 GHz
it contributes mainly to the signal in the high-frequency channels.

The cold flares at the dewar rim function as a cold ground shield
to block emission from the Earth:
emission from the Earth must diffract over the curved flares
before reaching the antenna side lobes.
The antenna gain at the flare edges
varies from -45 dB in the forward direction to -75 dB in the far back lobes.
Diffraction over the flares provides additional attenuation.
We have measured the far-field beam pattern of
the assembled payload (including the flares) over $4\pi$ sr
Based on the measured back-lobe response, 
the Earth should contribute less than 1 mK to any channel.
Both the flight train and Earth signals vary as the gondola is tipped.
We will use the in-flight tip scans to verify the predicted signals.

Atmospheric emission at 35 km altitude
contributes approximately 1 mK at 10 GHz.
Below 10 GHz the spectrum is flat;
even if the absolute amplitude were unknown
the spectral curvature would contribute
less than 0.2 mK differential signal between channels.
To first order, we will model this differential signal
using models of atmospheric transmission
(Liebe 1981; Danese \& Partridge 1989).
We will confirm the model by correlating the sky-target difference spectra
with airmass
as the balloon changes altitude during flight.
Tip scans in which the dewar tilts 15\deg ~in either direction from the zenith
provide a second cross-check on atmospheric emission.

\medskip
\normalsize

\refitem
Bartlett, J.G., and Stebbins, A., 1991, ApJ, 371, 8

\refitem
Becker, R.~H., et al.\ 2001, AJ, 122, 2850

\refitem
Bennett, C.~L., et al.\ 2003, ApJ, in press

\refitem
Bersanelli, M., et al.\ 1994, ApJ, 424, 517

\refitem
Danese, L., and Partridge, R.B.\ 1989, ApJ, 342, 604

\refitem
Djorgovski, S.G., et al.\ 2001, ApJ, 560, L5

\refitem
Fan, X., et al.\ 2002, AJ, 123, 1247

\refitem
Fixsen, D.J., Mirel, P., Kogut, A., and Seiffert, M., 2002,
Rev. Sci. Inst, 73, 3659

\refitem
Gnedin, N.Y., and Ostriker, J.P.\ 1997, ApJ, 486, 581

\refitem
Haiman, Z., and Loeb, A.\ 1997, ApJ, 483, 21

\refitem
Kogut, A., et al.\, 2003, ApJ, in press

\refitem
Liddle, A.R. and Lyth, D.H. 1995, MNRAS, 273, 1177

\refitem
Liebe, H.J.\ 1981, Radio Science, 16, 1183

\refitem
Mather, J., et al., 1999, ApJ, 512, 511

\refitem
Oh, S.~P.\ 1999, ApJ, 527, 16

\refitem
Oh, S.~P.\ 2001, ApJ, 553, 499

\refitem
Tegmark, M., et al.\ 1997, ApJ, 474, 1

\end{document}